\documentclass[reprint,superscriptaddress,aps,longbibliography]{revtex4-2}
\usepackage[utf8]{inputenc}
\usepackage[T1]{fontenc}
\usepackage{amsmath}
\usepackage{amssymb}
\usepackage{graphicx}
\usepackage[capitalize, nameinlink]{cleveref}
\usepackage{subfigure}
\graphicspath{{figures/}}
\usepackage{adjustbox}
\usepackage{booktabs}
\usepackage{multirow}

\usepackage{color}
\usepackage{rotating}
\usepackage{ulem}
\usepackage{natbib}
\usepackage{soul}

\begin{document}

\title{Dynamics of metastable contact soliton dissipative exchange flows in one-dimensional ferromagnetic channels}

\author{Medhanie Estiphanos}
\author{Ezio Iacocca}
\affiliation{Center for Magnetism and Magnetic Nanostructures, University of Colorado Colorado Springs, Colorado Springs, CO 80918 USA}

\date{\today}

\begin{abstract}
Dissipative exchange flows (DEFs) are large-amplitude boundary value solutions of ferromagnetic channels. In their low-injection limit, DEFs reduce to spin superfluids.  However, in the strong injection limit, nonlinearities dominate close to the injection site and a soliton is formed; this solution has been termed a contact soliton dissipative exchange flow (CS-DEF).  Here, we numerically investigate CS-DEF solutions in a moderate injection regime and a finite injection width. We find a solution where two metastable solitons coexist in the injection region. This solution is metastable in the sense that any perturbation to the system will eject one of the solitons out of the injection region.  Moreover, soliton dynamics can be excited when two injection regions are separated by a certain distance. We find that the ensuing DEF between the solitons induces a steady-state dynamics in which metastable solitons are continually ejected and nucleated.  Furthermore, and depending on the relative signs of the spin injections, the soliton dynamics possess a particular handedness and frequency related to the spin transfer torque delivered by the DEF. Our results provide insights into the transport of spin current by DEFs - where the interaction between DEFs and solitons suggests a mechanism for detaching contact-solitons from the injection boundary. Although this study focuses on the “nonlocal” interaction between solitons, it may lead to the investigation of new mechanisms for inserting solitons in a DEF, e.g., for discrete motion and transport of information over long distances.
\end{abstract}
\maketitle

\section{Introduction}

Long-distance spin transport has been theoretically studied~\cite{Konig2001,sonin2010spin,takei2014superfluid,Chen2014,takei2015nonlocal,Sonin2017,iacocca2017symmetry,Ochoa2018,Sonin2019,iacocca2019hydrodynamic,Zarzuela2020} because of the fundamental interest in stabilizing novel dynamical magnetization states. Although spin waves can achieve relatively long propagation distances~\cite{Madami2011,chumak2015magnon,Lebrun2018}, their spatial decay is exponential due to the effect of magnetic damping. In contrast, theoretical work has demonstrated that a large-amplitude chiral magnetization state in one-dimensional ferromagnets exhibits an algebraic spatial decay. This state has been termed spin superfluid~\cite{sonin2010spin,takei2014superfluid} or dissipative exchange flow (DEF)~\cite{iacocca2017symmetry,iacocca2019hydrodynamic}. The algebraic decay originates from the fact that the state results from a boundary value problem, and so it necessarily needs to span the full length of the channel. Physically, spin injection at one edge of the ferromagnetic channel induces an out-of-plane tilt in the magnetization which precesses due to shape anisotropy (local dipole for thin films) and propagates along the channel. Damping is then compensated nonlocally by spin injection, resulting in a chiral state with a low precessional frequency, on the order of MHz for typical metallic ferromagnets.

Because of the nonlocal compensation of damping, the stabilization of DEFs requires a large torque applied at the boundary. From the point of view of spin-transfer torque~\cite{slonczewski1996current}, this leads to prohibitively large currents~\cite{iacocca2019hydrodynamic,Estiphanos2024}. This is perhaps one of the main limitations that has hindered clear experimental observation of DEFs despite efforts in several materials~\cite{wesenberg2017long,Stepanov2018,Yuan2018,Lebrun2018}. However, it remains possible that alternative geometries, material systems, and spin injection mechanism can realize the necessary conditions to stabilize DEFs.

The chirality of DEFs also brings about interesting topological effects. For example, the dispersive hydrodynamic interpretation of the equations of motion~\cite{iacocca2017breaking} leads to the definition of a sonic curve. Therefore, upon interaction with an obstacle, the chirality of the DEF can be broken by vortex-antivortex pairs~\cite{iacocca2017vortex,Iacocca2020} akin to turbulence. This is also known as phase slips~\cite{Kim2016} and the underlying fluid state has been investigated as a mechanism for spin transport via defects~\cite{Kim2015} or thermal magnons~\cite{Flebus2016}. In addition, it is possible to develop solitons at the injection site. These have been termed contact solitons dissipative exchange flows (CS-DEFs)~\cite{iacocca2019hydrodynamic} or soliton screened spin transport~\cite{Schneider2021}. The profile of the contact solitons depends on both the injection site and the environment. For this reason, the problem can be approached from the point of view of boundary layers, in which the injection is dominated by exchange and the remainder of the channel is dominated by dissipation~\cite{iacocca2019hydrodynamic}. Recently, this property was used to modulate the soliton profile in time and therefore pump magnons into the channel~\cite{Estiphanos2024}.

Here, we investigate dissipative exchange flow solutions in a ferromagnetic channel with two injection sites with a finite width $w$ separated by a distance $L$. This ``well'' stabilizes either a linear DEF or a static uniform hydrodynamic state which is is exactly analogous to a superfluid~\cite{Konig2001}. In addition, we investigate a regime in which the injection is sufficient to stabilize CS-DEF solutions but is insufficient to completely reverse the magnetization under the injection region. The balance between exchange and local demagnetization (shape anisotropy) leads to metastable solutions composed of two solitons under the injection region. Coupling between the two injection sites is achieved by spin transport, destabilizing the solitons and finally reaching a steady-state regime of soliton dynamics. This result is of fundamental interest towards a better understanding of the interaction between spin transport and solitons. We also argue that our results illustrate how dynamic phenomena could be used to indirectly demonstrate the existence of spin hydrodynamics experimentally.

The remainder of the paper is organized as follows. In Sec. II, we describe the linear DEF solutions within the injection sites as a function of the relative sign of the fluid velocity boundaries. The analytical results are verified by use of a pseudospectral Landau-Lifshitz model~\cite{Rockwell2024} that provides a better resolution of the exchange energy and thus a more accurate model of soliton profiles. The metastable CS-DEF solutions are discussed in Sec. III for all the possible relative spin injection signs and we discuss how the steady-state solution is determined by spin-transfer torque carried by spin currents. The effect of initial conditions is also discussed. Finally, we provide our concluding remarks in Sec. IV.

\section{Confined hydrodynamic solutions}

The dynamics of the ferromagnetic channel are given by the Landau-Lifshitz equation
\begin{equation}
\label{eq:ll}
    \frac{\partial}{\partial t}\mathbf{m}=-\mathbf{m}\times\mathbf{H}_\mathrm{eff}-\alpha\mathbf{m}\left(\mathbf{m}\times\mathbf{H}_\mathrm{eff}\right),
\end{equation}
given here in a dimensionless form whereby $x\rightarrow x/\lambda_\mathrm{ex}$, with $\lambda_\mathrm{ex}$ is the exchange length, and $t\rightarrow \gamma\mu_0M_st$, where $\gamma$ is the gyromagnetic ratio, $\mu_0$ is the vacuum permeability, and $M_s$ is the saturation magnetization. The first term in Eq.~\eqref{eq:ll} is the Larmor torque equation and the second is the dissipative term scaled by the Gilbert damping constant, which is valid if $\alpha\ll1$. The normalized magnetization vector $\mathbf{m}=(m_x,m_y,m_z)$ maintains its norm in time and space. The effective field is considered to be $\mathbf{H}_\mathrm{eff}=\Delta\mathbf{m}-m_z\hat{z}$, where the first term is ferromagnetic exchange and the second is a negative uniaxial anisotropy that approximates the demag field for thin films. In other words, this effective field indicates that the modeled material is an in-plane ferromagnet.

Equation~\eqref{eq:ll} can be rewritten in a fluid-like form by invoking the transformation $n=m_z$ and $\mathbf{u}=-\nabla\left[\arctan(m_y/m_x)\right]$, where $n$ is the fluid density and $\mathbf{u}$ is the fluid velocity~\cite{iacocca2017breaking,Iacocca2019_Rev}. It is important to note that the gradient in the definition of $\mathbf{u}$ follows the natural Cartesian coordinate system. This implies that $\mathbf{u}>0$ if it oriented along $+\hat{x}$ while $\mathbf{u}<0$ if it is oriented along $-\hat{x}$. For a one-dimensional ferromagnetic channel, $\mathbf{u}=u\hat{x}$ and so we can consider the velocity as a signed scalar, $u$. The resulting dispersive hydrodynamic equations read
\begin{subequations}
\label{eq:fluid}
\begin{eqnarray}
\label{eq:fluid_n}
\partial_tn &=& \partial_x[(1-n^{2})u] +\alpha(1-n^{2})\partial_t\Phi,\\
\label{eq:fluid_u}
\partial_tu &=& -\partial_x[(1+u^{2})n] - \left[\frac{\partial_{xx}n}{{1-n^{2}}}+\frac{(\partial_xn)^2}{(1-n^{2})^2}\right]\nonumber\\ && +\alpha\partial_x\left[\frac{1}{1-n^{2}}\partial_x[(1-n^{2})u]\right].
\end{eqnarray}
\end{subequations}

\begin{figure}
    \centering
    \includegraphics[width=1\linewidth]{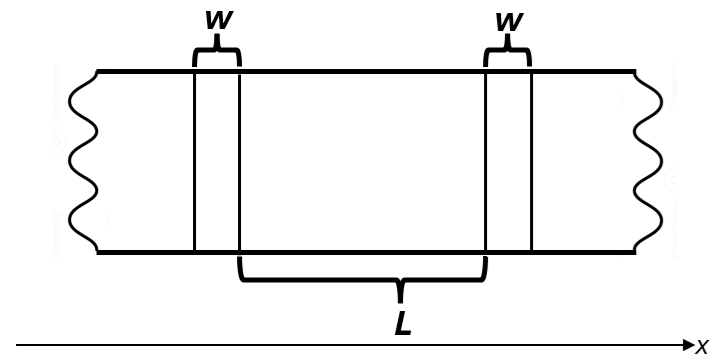}
    \caption{Schematic illustration of the central region of the one-dimensional ferromagnetic channel. The injection regions have identical widths $w$ and are separated by a distance $L$ measured from center-to-center.}
    \label{fig:schematic}
\end{figure}

We consider a long channel subject to spin injection in two regions of width $w$, as shown in Fig.~\ref{fig:schematic}. These regions are separated by a ``well'' distance $L$. For simplicity, we set the origin of our reference frame midway between the injection sites. Therefore, the region within the injectors can be approximately treated as a boundary value problem subject to boundary conditions
\begin{subequations}
    \label{eq:bc}
    \begin{eqnarray}
    \label{eq:bcn}
    \partial_tn(x = -L/2)=0,&\quad& \partial_tn(x = L/2)=0,\\
    \label{eq:bcu}
    u(x = -L/2)=\bar{u}_{L},&\quad& u(x=L/2)=\bar{u}_{R},
    \end{eqnarray}
\end{subequations}
where $\bar{u}_{L}$ and $\bar{u}_{R}$ represent the left and right fluid velocities~\cite{iacocca2019hydrodynamic}.

Let us consider the weak-injection regime, i.e., $0 < |\bar{u}| \ll min(1, \alpha L)$ to obtain a linear dissipative exchange flow (lDEF)~\cite{iacocca2017symmetry,iacocca2019hydrodynamic} which corresponds to the spin superfluid solution~\cite{sonin2010spin,takei2014superfluid}. In this case, the nonlinear terms in Eqs.~\eqref{eq:fluid} are negligible, we assume that $\partial_tn=0$ [immediately satisfying Eq.~\eqref{eq:bcn}], and we define $\Omega = \partial_{t}\phi$. The resulting equations are
\begin{subequations}
    \label{eq:linear}
    \begin{eqnarray}
    \label{eq:linearu}
    \alpha\Omega &=& -\partial_xu,\\
    \label{eq:linearn}
    \Omega &=& -n,
\end{eqnarray}
\end{subequations}
from which the solution for $u$ is obtained by simple integration
\begin{equation}
    \label{eq:sol_linear}
	u = -\alpha\Omega x + C,
\end{equation}
with $C$ as a constant of integration. Using the boundary conditions Eq.~\eqref{eq:bcu}, the constant of integration and precessional frequency can be uniquely found, yielding
\begin{subequations}
    \label{eq:solutionSSL}
    \begin{eqnarray}
    \label{eq:SSL}
u(x) &=& \bar{u}_{L}\left(\frac{1}{2} - \frac{x}{L}\right)  + \bar{u}_{R}\left(\frac{1}{2} + \frac{x}{L}\right),\\
    \label{eq:Omega}
    \Omega &=& -n = \frac{\bar{u}_{L}  - \bar{u}_{R}}{\alpha L}	
\end{eqnarray}
\end{subequations}

While Eq.~\eqref{eq:SSL} is general, it is interesting to study the cases where the injection magnitudes are the same on both sides. This gives rise to two different scenarios:

\subsection{Different signs: $\bar{u} = \bar{u}_{L} = -\bar{u}_{R}$}
\label{sec:diff}

This substitution yields 
\begin{eqnarray}
\label{eq:diffu}
u(x) &=& \bar{u}\left(\frac{1}{2} - \frac{x}{L}\right)  - \bar{u}\left(\frac{1}{2} + \frac{x}{L}\right)\nonumber\\
     &=& -\bar{u}\frac{2x}{L}.
\end{eqnarray}

This solution has the same form as a linear DEF or spin superfluid. Here, Eq.~\eqref{eq:Omega} indicates that the spin density is $n=2\bar{u}/(\alpha L)$, i.e. finite and small. Note that the boundary conditions are defined based on the gradient of the phase in the Cartesian reference frame. This means that the left boundary follows the reference frame but the right boundary is against the reference frame. Therefore, the physical spin injection in this case, e.g. as provided by spin-transfer torque~\cite{iacocca2019hydrodynamic} has the same sign.

\subsection{Same signs: $\bar{u} = \bar{u}_{L} = \bar{u}_{R}$}
\label{sec:same}

In this case, the substitution yields
\begin{eqnarray}
\label{eq:sameu}
u(x) &=& \bar{u}\left(\frac{1}{2} - \frac{x}{L}\right)  + \bar{u}\left(\frac{1}{2} + \frac{x}{L}\right)\nonumber\\
     &=& \bar{u}.
\end{eqnarray}

Here, the solution is a uniform fluid velocity which corresponds to the uniform hydrodynamic state (UHS) identified in Ref.~\cite{iacocca2017breaking} as an ideal solution and its corresponding superfluid state identified in Ref.~\cite{Konig2001}. Such a solution is allowed because the boundaries enforce the same ``input'' and ``output'' fluid velocities. As in the previous case, the reference frame informs that opposite spin injection is necessary to achieve the uniform solution. More importantly, the spin density $n$ and the frequency $\Omega$ are zero as per Eq.~\eqref{eq:Omega}, which is precisely the ideal limit where there is no dissipation, i.e., no dynamics.
\begin{figure}[t]
    \centering
    \includegraphics[width=3in]{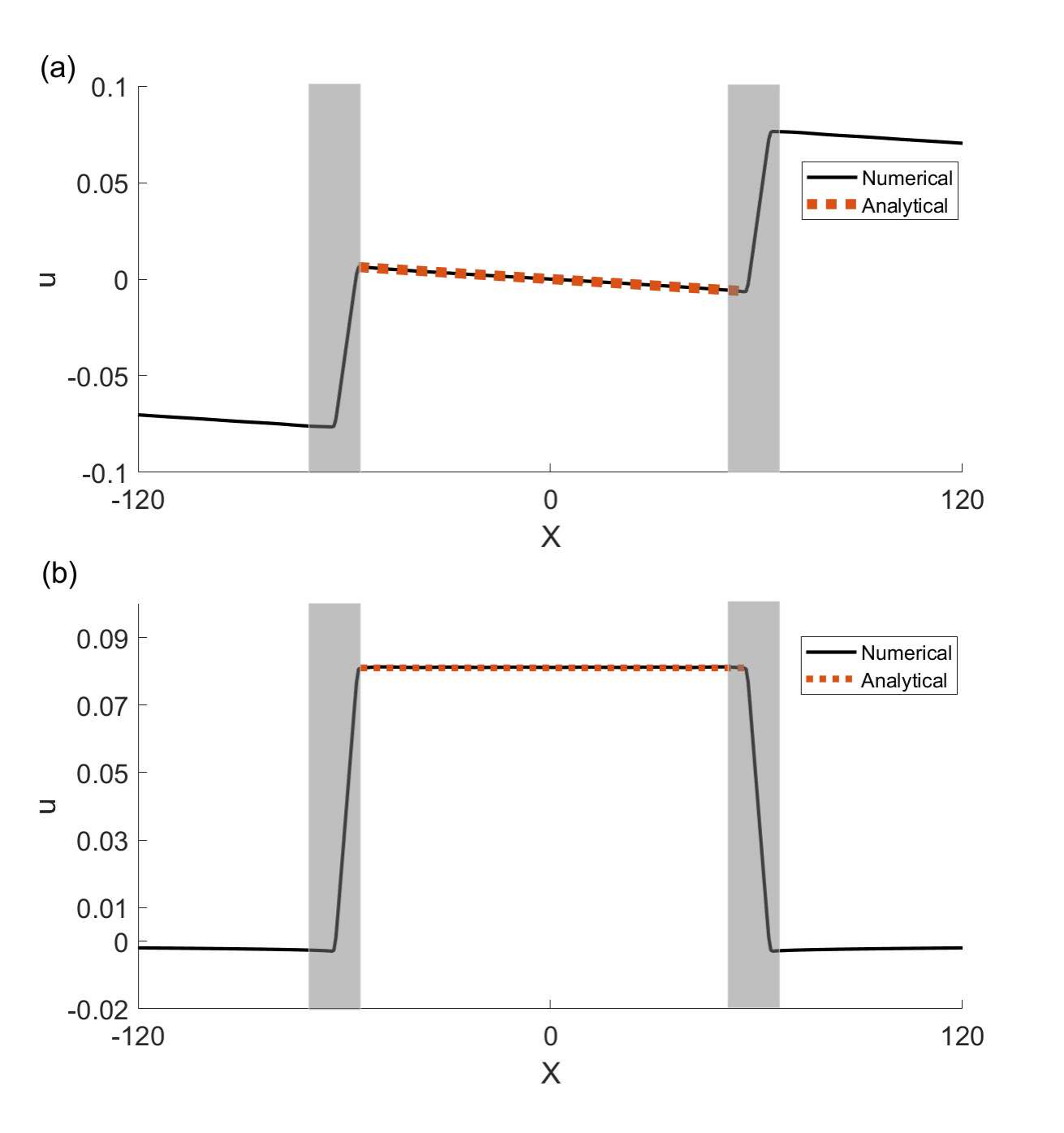}
    \caption{Fluid velocity computed in the weak injection regime, $\mu=0.01$. (a) When the spin injections have the same sign, a DEF is formed along the channel $L$. The analytical solution of Eq.~\ref{eq:diffu} upon use of a proportionality factor is overlaid with a red dashed line. (b) When the spin injections have different signs, a UHS is established along the channel. The analytical solution of Eq.~\ref{eq:sameu} is overlaid with a red dashed line, also using a proportionality factor. The gray area represent regions subject to spin injection.}
    \label{fig:linear}
\end{figure}

\subsection{Numerical implementation}

We verify the above solutions by numerical integration of the Landau-Lifshitz equation expressed in a pseudospectral form (PS-LL)~\cite{Rockwell2024}. In this implementation, the exchange term is replaced by a spectral kernel so that $\Delta\mathbf{m}\rightarrow\mathcal{F}^{-1}\{\omega(k)\hat{\mathbf{m}}(k)\}$, where $\mathcal{F}^{-1}\{\cdot\}$ represents the inverse Fourier transform, $\omega(k)=2-2\cos{(ka)}$ is the dispersion relation of magnons in dimensionless units and where $a$ is the lattice constant, and $\hat{\mathbf{m}}(k)$ is the Fourier transform of $\mathbf{m}$. This method has been useful to better describe the hydrodynamic solutions under strong injection where solitons are stabilized~\cite{Estiphanos2024} and which we study below.

In the PS-LL model, the spin injection is specified as a spin-transfer torque term~\cite{slonczewski1996current}. For simplicity, we collapse all physical parameters involved in spin transfer torque, i.e., current, spin polarization, and spin asymmetry, into the single constant $\bar{\mu}$. The spin injection $\bar{\mu}$ causes a tilt in the out-of-plane magnetization and it is thus proportional to $\bar{n}$, the fluid density at the injection site. There is a proportionality to the hydrodynamic injection $\bar{u}$, but this proportionality depends on the particularities of the channel as well as the geometry, i.e., injection width and length of the channel. We use magnetic parameters for permalloy in our simulations, namely: saturation magnetization $M_s=790$~kA/m, exchange length $\lambda_\mathrm{ex}=5$~nm, and Gilbert damping $\alpha=0.01$ as well as the gyromagnetic ratio $\gamma=28$~GHz/T and vacuum permeability $\mu_0=4\pi\times10^{-7}$~N/A$^2$. However, we report the results below in dimensionless units, where fields are scaled by $M_s$, space by $\lambda_\mathrm{ex}^{-1}$, and time by $\gamma\mu_0M_s\approx28$~GHz.

The linear regime is obtained for weak injection, e.g. $\bar{\mu}=\pm0.01$ applied in the left and right edges. In Fig.~\ref{fig:linear} we show results for (a) same spin injection signs and (b) different spin injection signs for a channel of length $800$, injection width $w=5.5$, and well distance $L=125$. The analytical solutions with appropriately scaled fluid velocities are overlaid with red dashed lines. In Fig.~\ref{fig:linear}(a), the same spin injection signs lead to a differently signed fluid boundary conditions and the solution is a DEF given by Eq.~\eqref{eq:diffu} where the proportionality factor was found to be $\approx0.817$. In Fig.~\ref{fig:linear}(b), the opposite spin injection signs lead to same-sign fluid boundary conditions and the solution is the UHS of Eq.~\ref{eq:sameu} and the proportionality factor was $\approx8.166$. The same qualitative solutions were obtained for a variety of well distances $L$, except for the proportionality factor between $\bar{\mu}$ and $\bar{u}$. It must be noted, however, that for the same spin injection signs, the proportionality factor depends on $L$ because $n=2\bar{u}/(\alpha L)$, as pointed out under Eq.~\eqref{eq:diffu} while for the case of different spin injection signs, the proportionality factor is approximately constant.
\begin{figure}[b]
    \centering
    \includegraphics[width=3.in]{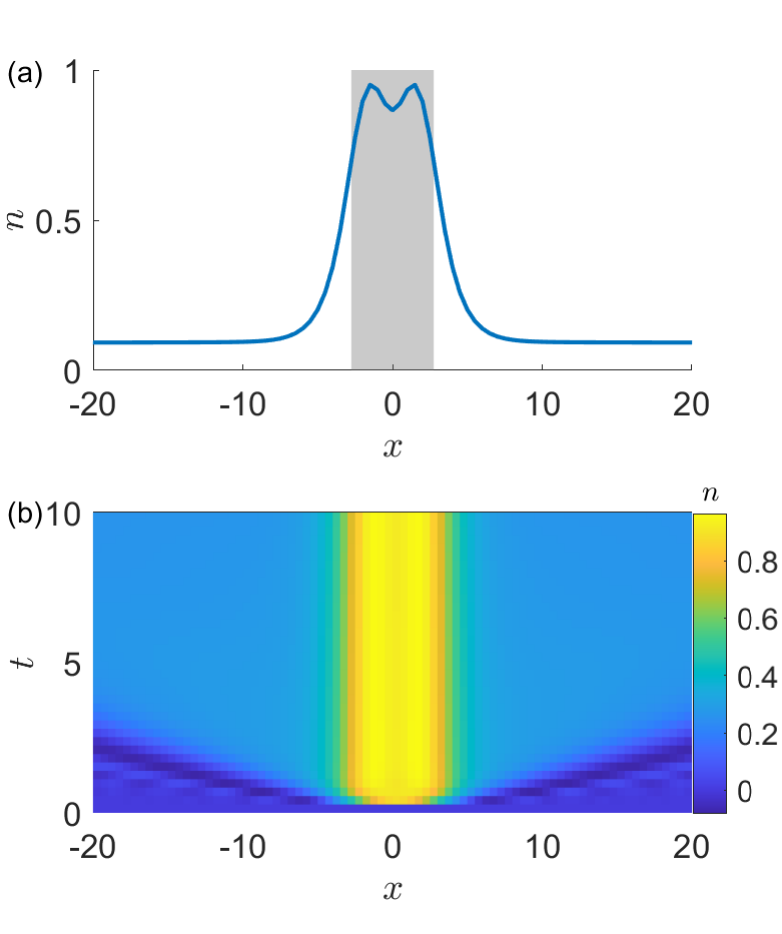}
    \caption{(a) Profile of two metastable solitons within the injection region, illustrated by a gray area. (b) Visualization of the metastable solitons in a spatiotemporal representation of $n$. The bright yellow contrast allows us to identify the position of the solitons in time.}
    \label{fig:soliton}
\end{figure}

\section{Metastable solitons}

We investigate the regime where the spin injection is strong enough to stabilize localized solutions under the injection region $w$. This solution has been termed contact soliton dissipative exchange flow or CS-DEF~\cite{iacocca2019hydrodynamic}. In a previous publication~\cite{Estiphanos2024}, the authors demonstrated that the CS-DEF can be forced to its extreme with spin injections on the order of $|\bar{\mu}|=1$. In this case, the injection region is fully reversed and the soliton is established in the channel with a width of approximately 0.66. Here, we are interested in an injection on the order of $|\bar{\mu}|=0.1$, in which CS-DEFs are established but the amplitude of the soliton is less than 1 resulting in the nucleation of two coexisting solitons.
\begin{figure}[t]
    \centering
    \includegraphics[width=3.3in, trim={0 1in 0 0.5in},clip]{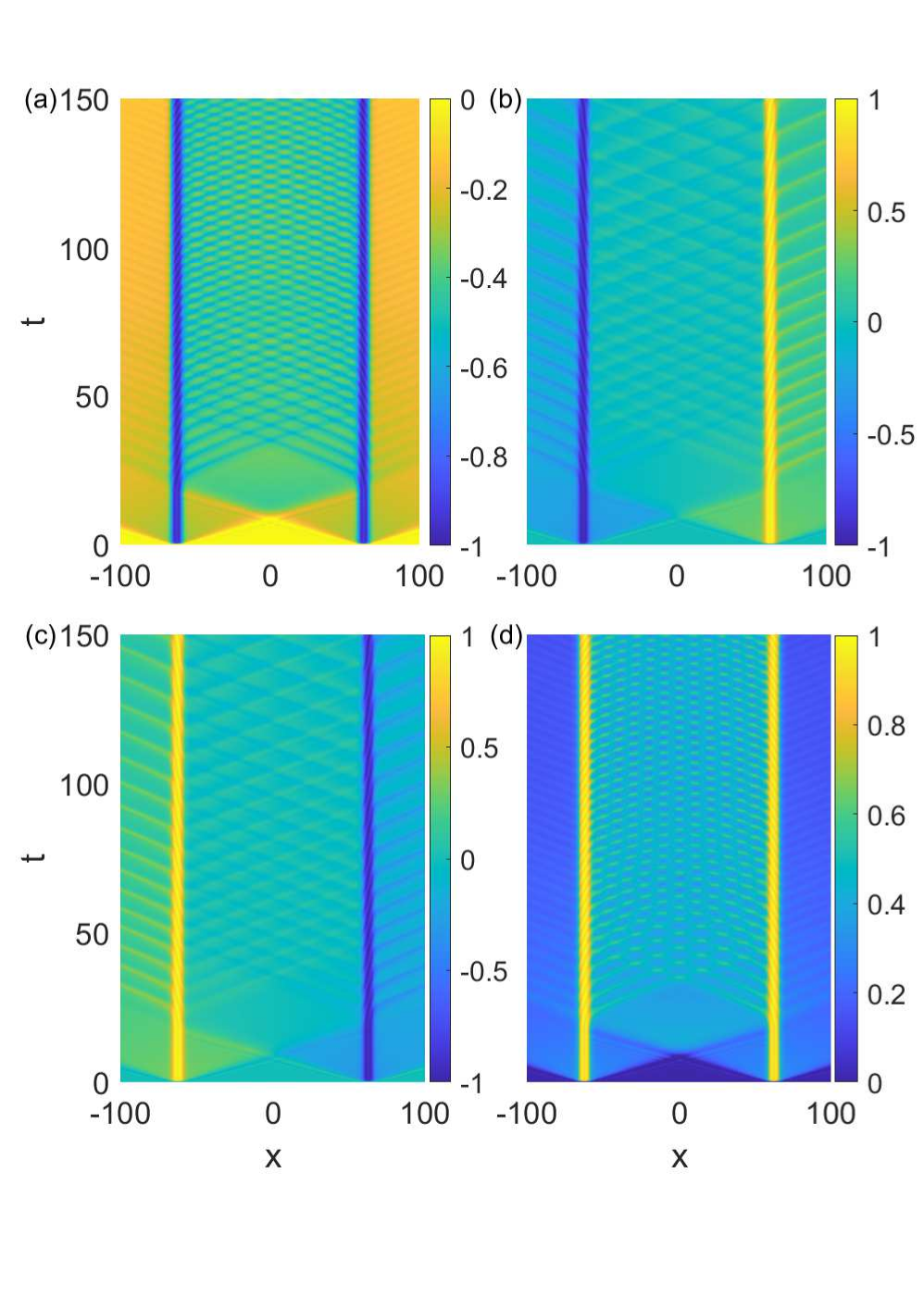}
    \caption{Metastable soliton interaction mediated by DEF in the cases (a) $\bar{\mu}_L=+0.1$, $\bar{\mu}_R=+0.1$, (b) $\bar{\mu}_L=0.1$, $\bar{\mu}_R=-0.1$, (c) $\bar{\mu}_L=-0.1$, $\bar{\mu}_R=+0.1$, and (d) $\bar{\mu}_L=-0.1$, $\bar{\mu}_R=-0.1$. The solitons are destabilized in all cases and develop an internal dynamical structure that settles into a steady state for $t>100$.}
    \label{fig:RelMu}
\end{figure}

A representative example is shown in Fig.~\ref{fig:soliton}(a), obtained for the same width $w$ and $\bar{\mu}=-0.1$. Clearly, two solitons can be stabilized under the region $w$. These solitons smoothly appear from the moment the injection is introduced. This can be visualized by the color plot of the spin density $n$ shown in Fig.~\ref{fig:soliton}(b). This internal structure can be only stabilized by a delicate balance between exchange and nonlinearity. In this case, the observed stability is a consequence of the symmetric simulation. It is important to note that the contact soliton solution ensures that the long-range state is a DEF. This means that the analytical solutions discussed in Sec.~\ref{sec:diff} and Sec.~\ref{sec:same} are approximately valid.

Returning to our geometry with two injection sites, the symmetry is broken and we expect the solitons to be unstable. The solutions are visualized as colorplots in Fig.~\ref{fig:RelMu} for the four possible combinations of spin injection: (a) $\bar{\mu}_L=+0.1$, $\bar{\mu}_R=+0.1$, (b) $\bar{\mu}_L=+0.1$, $\bar{\mu}_R=-0.1$, (c) $\bar{\mu}_L=-0.1$, $\bar{\mu}_R=+0.1$, and (d) $\bar{\mu}_L=-0.1$, $\bar{\mu}_R=-0.1$. The soliton destabilization is correlated to the development of a steady state within the well as well as an internal structure under the injection area. The steady state characteristics are dependent on the relative sign of the injections. In particular, Figs.~\ref{fig:RelMu}(a) and (d) have the same relative sign, corresponding to the situation where the fluid boundary conditions have different sign, section~\ref{sec:diff}. It is observed that when the solitons interact via DEFs, the internal soliton structure is destabilized. As a consequence, the internal solitons are ejected and decay along the well. This destabilization engenders internal dynamics under the injection region where solitons continually translate into the well. The dynamics continue until a steady state is achieved for $t>100$. This steady state implies that the internal dynamics are synchronized to the time required for the soliton to propagate along the well distance. For the case when the relative injection sign is different, Figs.~\ref{fig:RelMu}(b) and (c), the fluid boundaries correspond to the same sign, section~\ref{sec:same}. While a similar destabilization of the solitons is observed, the solitons are ejected outside of the well in this case. Similarly, a steady state is achieved at $t>100$.
\begin{figure}[b]
    \centering
    \includegraphics[width=3.3in]{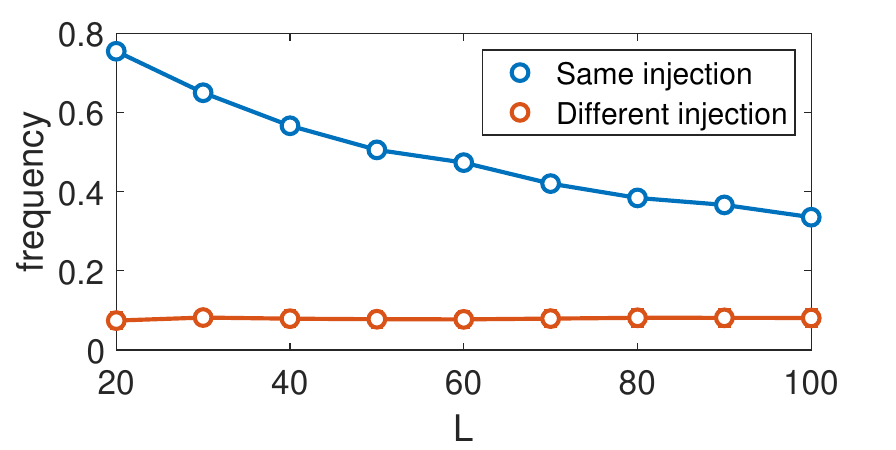}
    \caption{Frequency dependence of the steady state within the channel as a function of $L$.}
    \label{fig:freqs}
\end{figure}

We further characterize the steady state dynamics by investigating their frequency dependence on the well distance $L$. It was argued above that the internal dynamics are the result of the synchronization between the soliton ejection from the internal region and its translation through the well. To assess this statement, we plot the spectrum of the fluid density $n=m_z$ as a function of $L$. In particular, we run simulations for a time $t=556$ and collect the time-traces of $n$ within the well for $139<t<556$ discretized in $0.278$, from which we obtain a frequency resolution of $0.002$ and maximum frequency $1.798$. For the parameters of permalloy, this corresponds to a frequency resolution of $67$~MHz and maximum frequency $50$~GHz. To accumulate statistics, we find the frequency of the fundamental peak for time-traces at each spatial position in the well by fitting two harmonic Lorentzian functions. This is justified because the dynamics are expected to be coherent, although they present a significant harmonic content because of the discrete soliton ejections. The frequencies for the cases when the relative spin injection is the same are shown in Fig.~\ref{fig:freqs} by a blue curve and symbols. The frequency clearly decreases as a function of width. Interestingly, this decrease has a $1/L$ dependence similar to Eq.~\eqref{eq:diffu}. This can be intuitively understood as the solitons are ``carried'' by the underlying fluid flow. The case of different spin injection is shown in Fig.~\ref{fig:freqs} by a red curve and symbols. In contrast to the case of same singed injections, the frequency inside the well is approximately constant $0.07$ and does not change within errorbars. This is again understood because a UHS is stabilized in the channel and so the fluid flow is independent of the channel length $L$.
\begin{figure}[t]
    \centering
    \includegraphics[width=3.3in]{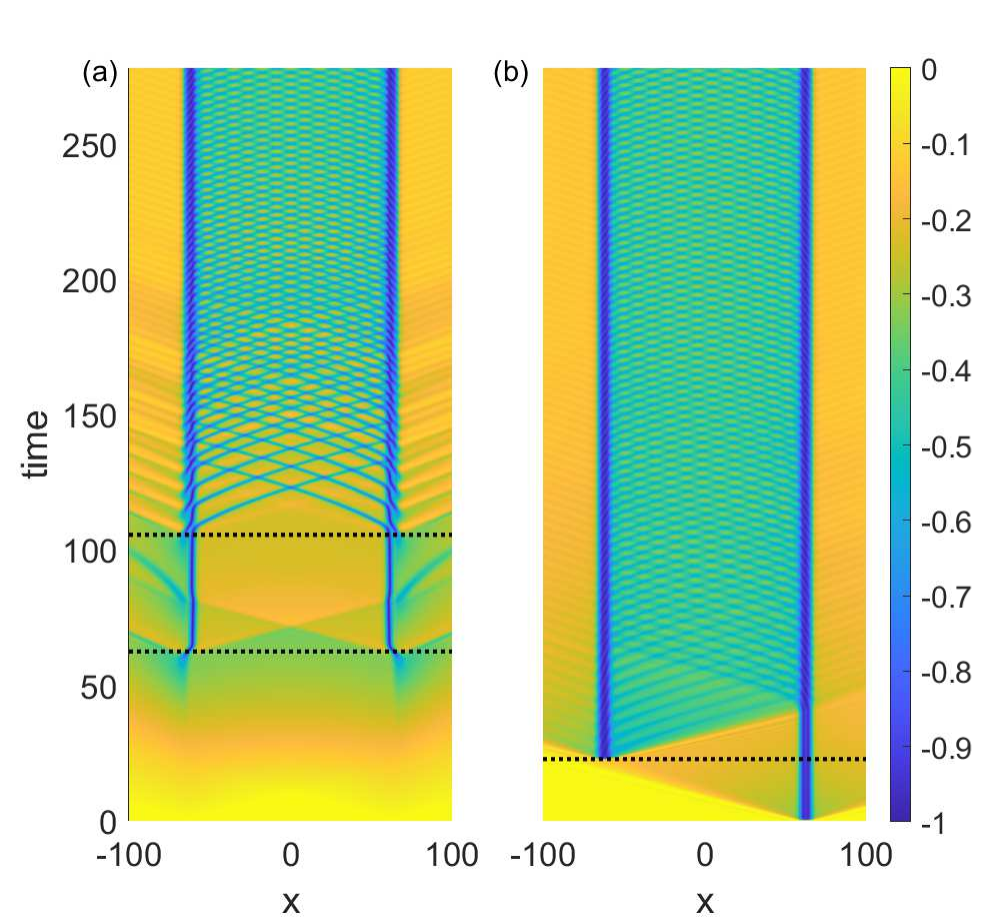}
    \caption{(a) Spatiotemporal evolution of $n$ when the current is linearly ramped from $\bar{\mu}=0$ at $t=0$ to $\bar{\mu}=0.1$ at $t=280$. A soliton is nucleated at $t\approx62.5$ and internal dynamics are not observed. At $t\approx105.6$ a second soliton is nucleated and the internal dynamics immediately ensue. (b) Spatiotemporal evolution of $n$ where $\bar{\mu}=0.1$ is injected in the right contact, but is delayed in the left contact by $t\approx22.8$. The dynamics also immediately ensue in the left contact and start in the right contact when the spin current reaches the metastable solitons.}
    \label{fig:tests}
\end{figure}

We return to the internal soliton dynamics. We distinguish the ``handedness'' of such dynamics as being dependent on the relative injection sign. Because the handedness depends solely on the relative injection sign, it must be related to the nature of the initial interaction between the contact solitons, i.e., for $t<\approx 25$ in Fig.~\ref{fig:RelMu}. This interaction is characterized by a dispersive shock wave~\cite{El2016,Hu2021,Hu2022} which arises due to the instantaneous spin injection. The dispersive shock wave is, in essence, the leading structure of the DEF and therefore carries the spin basis of the DEF proportional to the spin injection. As a consequence, we can consider that the dispersive shock wave and DEF carry a spin current. When the spin current impinges on the soliton, it results in a torque that arises due to conservation of angular momentum~\cite{slonczewski1996current}. In fact, this scenario is similar to the Zhang-Li torque~\cite{Zhang2004} and magnonic torques~\cite{Ai2023} which are known to cause domain-wall motion.

Based on the above description, we can consider two cases. When the relative spin injections have the same sign, the solitons and the spin bases have the same sign. By conservation of angular momentum, as the spin current passes through the soliton, the local magnetization aligns with the spin current resulting in a net motion of the soliton into the well. The second case occurs when the relative spin injection sign is different. Here, conservation of angular momentum indicates that the spins attempt to reverse, resulting in a net motion of the soliton out of the well. It is important to note these dynamics are possible because the injection has a finite width and the region supports two metastable solitons.
 
It is natural to inquire if the soliton dynamics require phase coherence or even the presence of a dispersive shock wave. To address this question, we performed simulations when the spin injection magnitude is ramped up linearly for $t=280$ until it reaches the nominal injection value, upon which is then held constant. We have focused only on the case of same spin injection sign. The spatiotemporal evolution of $n$ is shown in Fig.~\ref{fig:tests}(a) In this case, there is no visible dispersive shock wave interacting with the soliton. In fact, DEF solutions are established as the injection continues to grow. At $t\approx62.5$, a soliton is nucleated and a shock is produced from the event. However, it is only at $t\approx105.6$ where two solitons appear and the internal dynamics ensue. This points to the fact that the transient dynamics do not play a significant role in the soliton dynamics but rather the spin current carried by the DEF. Notice that the soliton dynamics occur at a much lower spin injection than explored earlier. In fact, from this simulation we can indicate that a soliton is nucleated when the spin injection magnitude is $\bar{\mu}\approx0.02$ and two metastable solitons when $\bar{\mu}\approx0.04$.

To further corroborate the above statements, we perform simulations where the spin injection is enabled at different and random times. One example is shown in Fig.~\ref{fig:tests}(b) where the leftmost spin current starts at $t\approx22.8$. Clearly, the internal dynamics appear immediately, regardless of the fact that the initial dispersive shock wave is already past the spin injection location. Conversely, the rightmost solitons are destabilized when subject to the the dispersive shock wave originating from the leftmost injection. This is proof that the DEF carries the angular momentum needed to initiate the internal dynamics.

\section{Conclusions}

We have investigated the dynamics of metastable solitons subject to a spin current carried by a DEF. Metastable solitons are present in a nanowire at moderate injection strengths, approximately an order of magnitude lower than those needed to nucleate well-defined solitons~\cite{Estiphanos2024}. Several features of spin hydrodynamics are demonstrated in this geometry. First, we show that a uniform hydrodynamic state, which is a truly dissipationless spin fluid, can be established by utilizing the same injection on both edges. This state immediately dissipates energy as it becomes dynamic but we show that it supports fluctuations, i.e., soliton propagation.

The internal soliton dynamics are studied in more detail, specifically the role of spin current mediated by the dissipative exchange flow, which can result in metastable solitons being destabilized by spin transfer torque and eventually dislodged from the contact region. This could be further investigated as a means for encoding information on a spin hydrodynamic state. For example, future research could involve the positioning of a soliton as a defect on a spin hydrodynamic state to enable discrete propagation of information. From the perspective of an experimental demonstration of spin hydrodynamics, the suggested geometry of a nanowire with two contacts providing spin-transfer or spin-orbit torques is feasible with state-of-the-art technology. The dynamics within the well could then be probed by a third contact region by e.g., inverse spin Hall effect or optically with Brillouin light scattering. The different homogeneous frequency depending on the relative spin injection signs would provide an unambiguous verification of the stabilization of a spin hydrodynamic state. In addition, our results provide a deeper insight into the fluid-like behavior of magnets, particularly in the nonlinear limit where magnetic solitons exist and interact with spin currents.

\section*{Acknowledgments}

This work was supported by the U.S. Department of Energy, Office of Basic Energy Sciences under Award Number DE-SC0024339.

\end{document}